\documentclass[12pt]{article}
 
\usepackage{epsfig}
\begin{document}

\baselineskip=14pt plus 0.2pt minus 0.2pt
\lineskip=14pt plus 0.2pt minus 0.2pt

%***********************
\newcommand{\be}{\begin{equation}}
\newcommand{\ee}{\end{equation}}
\newcommand{\da}{\dagger}
\newcommand{\dg}[1]{\mbox{${#1}^{\dagger}$}}
\newcommand{\hlf}{\mbox{$1\over2$}}
\newcommand{\lfrac}[2]{\mbox{${#1}\over{#2}$}}
%**********************

\begin{flushright}
quant-ph/9908048 \\
LA-UR-99-1157 \\
\end{flushright} 

\begin{center}
\large{\bf Higher-Power Coherent and Squeezed States
\footnote{\noindent  Dedicated to Marlan Scully on he occasion of his 
60h birthday.}}
 
\vspace{0.25in}

%\large
\bigskip

Michael Martin Nieto\footnote{\noindent  Email:  
mmn@pion.lanl.gov}\\
{\it Theoretical Division (MS-B285), Los Alamos National Laboratory\\
University of California\\
Los Alamos, New Mexico 87545, U.S.A. \\}
and\\
{\it Universit\"at Ulm, 
Abteilung f\"ur Quantenphysik, \\
Albert-Einstein-Allee 11, 
D 89069 Ulm, 
GERMANY}\\

\vspace{0.25in}

 D. Rodney Truax\footnote{Email:  truax@acs.ucalgary.ca}\\
{\it Department of Chemistry\\
 University of Calgary\\
Calgary, Alberta T2N 1N4, Canada\\}
 
\normalsize

%***********************************************
\vskip 20pt
\today
%**************************************************

\vspace{0.3in}

{ABSTRACT}
 
\end{center}
\begin{quotation}
%********************************************************************
\baselineskip=.33in
%******************************************************************
A closed form expression for the  higher-power coherent states 
(eigenstates of $a^{j}$) is given.  The cases $j=3,4$ are discussed in 
detail, including the time-evolution of the probability densities.  
These are compared to the case $j=2$, the even- and odd-coherent 
states.  We give the extensions to the ``effective" displacement-operator, 
higher-power squeezed states and to the
ladder-operator/minimum-uncertainty, higher-power squeezed states.
The properties of all these states are discussed.

\vspace{0.25in}


\end{quotation}

\newpage

%********************************************************************
\baselineskip=.33in
%******************************************************************

%********************1. HPCS**********************

\section{Definitions of Higher-Power Coherent States}

Higher-power coherent states (HPCS) 
are defined by  \cite{manko}-\cite{nagel2}
\be
a^j |\alpha;j,k\rangle = \alpha^j |\alpha;j,k\rangle, ~~~~~~~
    0 \le k \le (j-1),   \label{hpcs}
\ee
\be
\alpha = \alpha_1 +i\alpha_2 \equiv \frac{x_0 +ip_0}{\sqrt{2}}.
\ee
This definition produces ladder-operator-type 
coherent states (LOCS), 

A second type of coherent states, 
generally equivalent to the LOCS, are minimum-uncertainty 
coherent states (MUCS).  These come from  considering the operators
\begin{equation}
{X_j} = \frac{a^j + (a^{\dagger})^j}{\sqrt{2}},
         ~~~~~~~~~~{P_j} = \frac{a^j-(a^{\dagger})^j}{i\sqrt{2}}, 
\label{xp}
\end{equation}
with commutation relation
\be
[X_j,P_j] = i{\cal O},
\ee
${\cal O}$ being Hermitian. This implies a Heisenberg uncertainty relation
\begin{equation}
(\Delta {X_j})^2 (\Delta {P_j})^2 \ge \lfrac{1}{4}
               |\langle[X_j,P_j]\rangle|^2 
       = \lfrac{1}{4} \langle {\cal O} \rangle^2.   \label{hur}
\end{equation}
The (wave-function) states which satisfy equality in Eq. (\ref{hur}) are
given by solutions to the equation
\be
[X_j + iB P_j]\psi_{mu} = C \psi_{mu}, \label{mueq}
\ee
where 
\be
B = \frac{\langle{\cal O}  \rangle}{2(\Delta {P_j})^2} 
  = \frac{2(\Delta {X_j})^2}{\langle {\cal O} \rangle}, ~~~~~~~~~
C = \langle X_j \rangle + iB \langle P_j \rangle .  
\ee
These solutions, $\psi_{mu}$ comprise not only the coherent states 
but also some of the  squeezed states (SS) for the system.  
(See Sec. 5 below.  Remember, the 
CS are special-case, zero-squeezed SS.) 

To restrict the $\psi_{mu}$ to the $\psi_{cs}$, one needs to add the further 
restriction $\Delta {X_j}/\Delta {P_j}= Const$.  Given our 
present overall normalizations for $X_j$ and $P_j$, 
this constant is unity. For the harmonic oscillator, one 
has that the uncertainties in $x$ and $p$ are equal. 

For general potential systems,  this constant can be determined by 
the demand that the set of CS include the ground-state \cite{mucs}.
(The ground state is the quantum analogue of zero classical motion).    
However, here things are 
complicated by the fact that the HPCS have $j$ ``effective'' extremal states, 
not just the ground state.  Therefore, for a given $(j,k)$, 
each set  $ |\alpha;j,k\rangle$ does not span the Hilbert space. 

To continue, observe that there is no  displacement-operator 
coherent-state (DOCS) definition
\be
D_j(\alpha)|0\rangle =|\alpha;j,k\rangle, 
\ee
\begin{equation}
D_j(\alpha) = \exp\left[(\alpha^*)^j a^j - \alpha^j (a^{\dagger})^j\right], 
\end{equation}
for HPCS with $j> 2$.   
When $j> 2$, $a^{j}$ and $(a^{\da})^{j}$
do not form part of a closed algebra and  $D_j(\alpha)$
is not defined.   $\langle 0| D_{j>2}|0 \rangle$ 
does not converge in a power series evaluation \cite{nagel2, doss}.
(Even for $j=2$, potential definitions do not work.  See Sec. 2.)  

%*************************Properties of HPCS*******************

\section{Properties of Higher-Power Coherent States}

For $(j,k) = (1,0)$, the HPCS  are the ordinary coherent states.  
\be
|\alpha;1,0\rangle =\exp[-\lfrac{1}{2}|\alpha|^2] \sum_{n=0}^{\infty}
            \frac{\alpha^n}{\sqrt{n!}}|n\rangle 
\rightarrow \frac{\exp[-\lfrac{1}{2}(x-x_0)^2] 
             e^{i[p_0x - x_0 p_0]}}{\pi^{1/4}}. \label{cs}
\ee
(We use units $\hbar = m =\omega=1$.)
Up to a phase, the three LOCS, MUCS, and DOCS definitions yield the 
same Eq. (\ref{cs}).

\begin{figure}[ht]
 \begin{center}
\noindent    
\psfig{figure=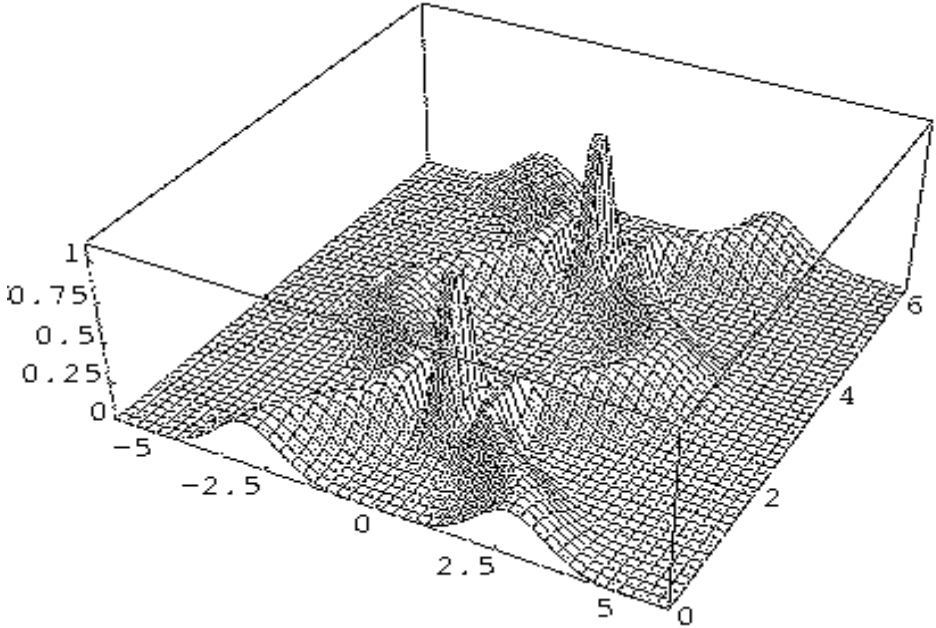,width=125mm,height=90mm}
  \caption{The time evolution of 
the even-coherent state $\rho_{(2,0)}(x,t)$ 
for the initial conditions $x_0 = 2^{3/2}$ and $p_0 = 0$.
 \label{fig:eocsfig1}}
 \end{center}
\end{figure}

\begin{figure}[ht]
 \begin{center}
\noindent    
\psfig{figure=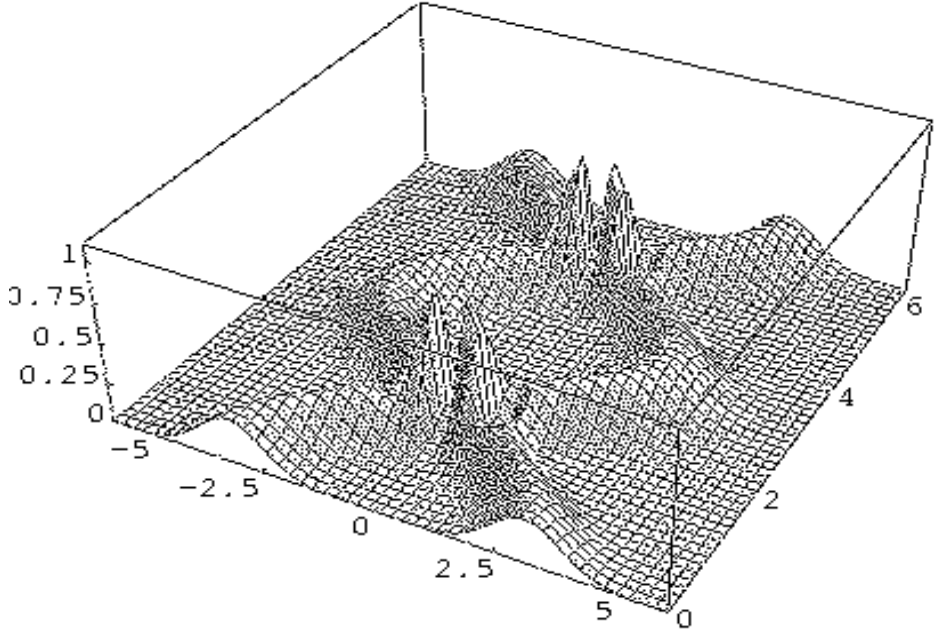,width=125mm,height=90mm}
  \caption{The time evolution of 
the odd-coherent state $\rho_{(2,1)}(x,t)$ 
for the initial conditions $x_0 = \sqrt{10}$ and $p_0 = 0$. 
 \label{fig:eocsfig2}}
 \end{center}
\end{figure}

The even- and odd-coherent states 
\cite{manko, vogel, eoss} are HPCS with $(j,k) = (2,0)$ and $(2,1)$. 
The LOCS and MUCS methods both lead to 
\begin{eqnarray}
|\alpha;2,0\rangle &=& [\cosh{|\alpha|^2}]^{-1/2}
      \sum_{n=0}^{\infty}\frac{\alpha^{2n}}{\sqrt{(2n)!}}|2n\rangle 
\rightarrow  \psi_{+}(x),  \\
|\alpha;2,1\rangle &=& [\sinh{|\alpha|^2}]^{-1/2}
\sum_{n=0}^{\infty}\frac{\alpha^{2n+1}}{\sqrt{(2n+1)!}}|2n+1\rangle 
\rightarrow  \psi_{-}(x).
\end{eqnarray}
\begin{equation}
\psi_{\pm}(x)
    = \frac{e^{-i2 x_0 p_0}\left[\exp[-\lfrac{1}{2}(x-x_0)^2] e^{i p_0 x}
	    \pm \exp[-\lfrac{1}{2}(x+x_0)^2]e^{ - i p_0 x}\right]}
      {2^{1/2} \pi^{1/4}\left[1 \pm \exp[-(x_0^2+p_0^2)])\right]^{1/2}}.
\label{eocs}
\end{equation} 
The wave packets of these  states are two Gaussians, at positions $\pi$ 
apart in the phase-space circle.  The even states are composed of 
$n=0,2,4,\dots$ number states.  These Gaussians, when they interfere, 
have a maximum central peak \cite{vogel,eoss}. (See Figure 
\ref{fig:eocsfig1}.) 
The odd states are composed of 
$n=1,3,5,\dots$ number states.   When the odd Gaussians interfere there is
a central minimum and two slightly smaller peaks on each side  
\cite{vogel,eoss}. (See Figure \ref{fig:eocsfig2}.)  

%************

%**************

%************

%**************

As stated above, the DOCS method does not work for $j>2$ and 
already has problems  
even for the j = 2 case.  One might think a ``viable'' displacement 
operator could be given by the form of the ordinary 
squeeze operator $S(z=2\alpha^2)$ of Eq. (\ref{opS}) below:
\be
D_2(\alpha) = \exp\left[(\alpha^*)^2 a^2 - \alpha^2 (a^{\dagger})^2\right].
\ee
Applying this operator to the true extremal state $|0\rangle$ does
produce an even state.  To obtain an odd state, you have to  
apply $S$ by hand to the state $|1\rangle$, which is outside the usual 
method.  However, the even and odd states so produced are not the 
even- and odd-CS.  Rather, they are the squeezed (but not displaced) 
number states with  $n=(0,1)$ \cite{sns}.  
One can also devise ``effective''  displacement operators 
\cite{manko}
\be
D_{\pm}(\alpha) = N_{\pm}[D_1(\alpha) \pm D_1(-\alpha)],   ~~~~~~
D_{\pm}(\alpha)|0\rangle = |\alpha;2, \lfrac{1}{2}\mp\lfrac{1}{2}\rangle.
\ee
But these operators are  not unitary: 
$ D_{\pm}(\alpha) D_{\pm}^{\dagger}(\alpha) \ne I$.

\indent From these $j=1,2$ examples, it is clear what occurs in general 
for higher-$j$ HPCS of the LOCS/MUCS variety.  
Given $j$ and $\alpha$, there are  $j$ states, labeled by $(j,k)$ with 
$0\le k \le (j-1)$.  They are separately composed of the 
number states: $\{ 0,j,2j, \dots \}$, $\{ 1,(j+1),(2j+1), \dots \}$, 
$\{ 2,(j+2),(2j+2), \dots \}$, $\dots$,  and $\{ (j-1),(2j-1),(3j-1), 
\dots\}$. 
In wave-function space, these $j$ states
all each contain $j$ Gaussians, separated by $2 \pi/j$ in the 
phase-space circle.  The relative phases among the $j$ Gaussians in 
each state are adjusted so that  the $(j,k)$-power states are 
mutually orthogonal.  

Using higher-order Hermite generating function techniques \cite{hermite}, 
one can use Eq. (\ref{hpcs}) to 
obtain closed-form expressions for these higher-power states. 
Specifically, the orthonormal states are 
\begin{eqnarray}
|\alpha;j,k\rangle &=&{\cal S}^{-1/2}(j,k,|\alpha|^2)
 \sum_{n=0}^{\infty}\frac{\alpha^{jn+k}}{\sqrt{(jn+k)!}}|jn+k\rangle 
      \\
&\rightarrow&\psi_{(j,k)}(x)  
= \frac{\exp\left[-\frac{1}{2}x^2\right]~G(j,k,x,\alpha/\sqrt2)}
     {\pi^{1/4}~ {\cal S}^{1/2}(j,k,|\alpha|^2)}.  \label{jkbasis}
\end{eqnarray}
In the above,  
${\cal S}$ is the sum
\begin{equation}
{\cal S}(j,k,z)=\sum_{n=0}^{\infty}\frac{z^{jn+k}}{(jn+k)!}
= \frac{1}{j}\sum_{l=1}^{j}
  \frac{\exp[z e^{i2\pi l/j}]}
       {[e^{i2\pi l/j}]^k},~ \label{Sjk}
\end{equation}
and $G$ is the higher-order Hermite generating function 
\cite{hermite}
\begin{equation} 
G(j,k,x,z) = \sum_{n=0}^{\infty}\frac{z^{jn+k}H_{jn+k}(x)}{(jn+k)!}
= \frac{1}{j}\sum_{l=1}^{j}
  \frac{\exp[-z^2 e^{i4\pi l/j}]\exp[2xz e^{i2\pi l/j}]}
       {[e^{i2\pi l/j}]^k}.  
    \label{Gjk}
\end{equation}

%**************************j=3 Special Case CS***********************

\section{Special-Case Higher-Power Coherent States}

\subsection{The Case $j=3$}

Let us now look at the $j=3$ states in detail.  First define
\begin{eqnarray}
A &=& \frac{x_0^2 + p_0^2}{2},  \label{A}  \\
N_{(3,0)} &=&\left[1+2\cos(\lfrac{\sqrt{3}}{2}A)\right],
     \label{N30} \\
N_{(3,1)} &=&\left[1-\left(\cos(\lfrac{\sqrt{3}}{2}A)
     -  \sin(\lfrac{\sqrt{3}}{2}A)\right)e^{-3A/2}\right],
     \label{N31} \\
N_{(3,2)} &=&\left[1-\left(\cos(\lfrac{\sqrt{3}}{2}A)
     +  \sin(\lfrac{\sqrt{3}}{2}A)\right)e^{-3A/2}\right],
     \label{N32}  \\
Y_1&=&\exp\left[-\lfrac{1}{2}[x  + (\lfrac{1}{2}x_0
                +\lfrac{\sqrt{3}}{2}p_0)]^{2}\right]           
               e^{i[x(\lfrac{\sqrt{3}}{2}x_0 -\lfrac{1}{2}p_0)
          +\lfrac{\sqrt{3}}{8}(x_0^2 - p_0^2) + \lfrac{1}{4}x_0p_0]},
    \label{Y1}  \\
Y_2&=&\exp\left[-\lfrac{1}{2}[x+(\lfrac{1}{2}x_0-\lfrac{\sqrt{3}}{2}p_0)]^{2}
              \right]
         e^{i[-x(\lfrac{\sqrt{3}}{2}x_0 +\lfrac{1}{2}p_0)
          +\lfrac{\sqrt{3}}{8}(x_0^2 - p_0^2) + \lfrac{1}{4}x_0p_0]}, 
     \label{Y2}  \\
Y_3&=&\exp\left[-\lfrac{1}{2}[x-x_0]^{2}\right]
         e^{i[xp_0 - \lfrac{1}{2}x_0p_0]},    
\label{Y3}  
\end{eqnarray}
The three, orthonormal, 3-power coherent states are then 
\begin{eqnarray}
\psi_{(3,0)}(x) &=& \frac{Y_1+Y_2 +Y_3}
               {3^{1/2} ~ \pi^{1/4} ~ N_{(3,0)}^{1/2}}
\label{psi30}   \\
\psi_{(3,1)}(x) &=& \frac{\left[-\lfrac{1}{2} -i\lfrac{\sqrt{3}}{2}\right]Y_1 
        +\left[-\lfrac{1}{2} +i\lfrac{\sqrt{3}}{2}\right]Y_2 +Y_3}
         {3^{1/2} ~ \pi^{1/4} ~ N_{(3,1)}^{1/2}}
\label{psi31}    \\
\psi_{(3,2)}(x) &=& \frac{\left[-\lfrac{1}{2} +i\lfrac{\sqrt{3}}{2}\right]Y_1 
+\left[-\lfrac{1}{2} -i\lfrac{\sqrt{3}}{2}\right]Y_2 +Y_3}
                     {3^{1/2} ~ \pi^{1/4} ~ N_{(3,2)}^{1/2}}
\label{psi32}     
\end{eqnarray}

If we define the angles, 
\begin{eqnarray}
\phi_{1,2} &=& x(\sqrt{3} x_0) + \lfrac{\sqrt{3}}{4}(x_0^2 - p_0^2), \\
\phi_{1,3} &=& x(\lfrac{\sqrt{3}}{2} x_0 -\lfrac{3}{2} p_0)
    + \lfrac{\sqrt{3}}{8}(x_0^2 - p_0^2) +\lfrac{3}{4}x_0 p_0, \\
\phi_{2,3} &=& x(\lfrac{\sqrt{3}}{2} x_0 +\lfrac{3}{2} p_0)
    - \lfrac{\sqrt{3}}{8}(x_0^2 - p_0^2) +\lfrac{3}{4}x_0 p_0,
\end{eqnarray}
then the three probability densities  
\be
\rho_{(3,k)}(x) = \psi_{(3,k)}^*(x) \psi_{(3,k)}(x) 
\ee
are
\begin{eqnarray}
\rho_{(3,0)}(x) &=& \frac{1}{3 \pi^{1/2} N_{(3,0)}}
           \left\{|Y_1^2| + |Y_2^2| + |Y_3^2|  
         + 2 \cos{\phi_{1,2}}~|Y_1 Y_2|  \right.  \nonumber \\
 &~&~\left. 
         + 2 \cos{\phi_{1,3}}~|Y_1 Y_3|
         + 2 \cos{\phi_{2,3}}~|Y_2 Y_3|\right\}, \\
\rho_{(3,1)}(x) &=& \frac{1}{3 \pi^{1/2} N_{(3,1)}}
           \left\{|Y_1^2| + |Y_2^2| + |Y_3^2|  \right.  
         - (\cos{\phi_{1,2}}+\sqrt{3}\sin{\phi_{1,2}})|Y_1 Y_2| \nonumber \\
 &~&~\left. 
         - (\cos{\phi_{1,3}}-\sqrt{3}\sin{\phi_{1,3}})|Y_1 Y_3|
         - (\cos{\phi_{2,3}}+\sqrt{3}\sin{\phi_{2,3}})|Y_2 Y_3|\right\}, \\
\rho_{(3,2)}(x) &=& \frac{1}{3 \pi^{1/2} N_{(3,2)}}
           \left\{|Y_1^2| + |Y_2^2| + |Y_3^2|  \right.  
         - (\cos{\phi_{1,2}}-\sqrt{3}\sin{\phi_{1,2}})|Y_1 Y_2| \nonumber \\
 &~&~\left. 
         - (\cos{\phi_{1,3}}+\sqrt{3}\sin{\phi_{1,3}})|Y_1 Y_3|
         - (\cos{\phi_{2,3}}-\sqrt{3}\sin{\phi_{2,3}})|Y_2 Y_3|\right\}.
\end{eqnarray}

Because we are working in an harmonic-oscillator system, time-dependence 
is achieved by taking
\be
x_0 \rightarrow x_0\cos{t} + p_0\sin{t}, ~~~~~~
p_0 \rightarrow p_0\cos{t} - x_0\sin{t}. \label{xtpt}
\ee 
in $\psi_{(3,k)}(x)$ and   $\rho_{(3,k)}(x)$.
 
\begin{figure}[ht]
 \begin{center}
\noindent    
\psfig{figure=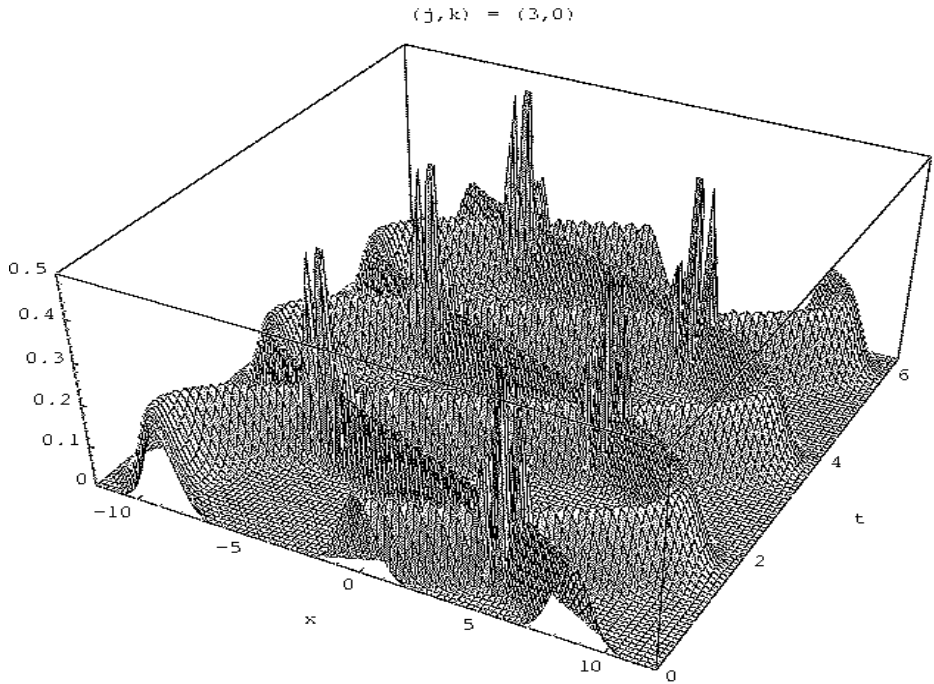,width=125mm,height=90mm}
  \caption{ The time evolution of $\rho_{(3,0)}(x,t)$ 
for the initial conditions $x_0 = 0$ and $p_0 = 10$. 
 \label{fig:hpcs3k=0}}
 \end{center}
\end{figure}

\begin{figure}[p]
 \begin{center}
\noindent    
\psfig{figure=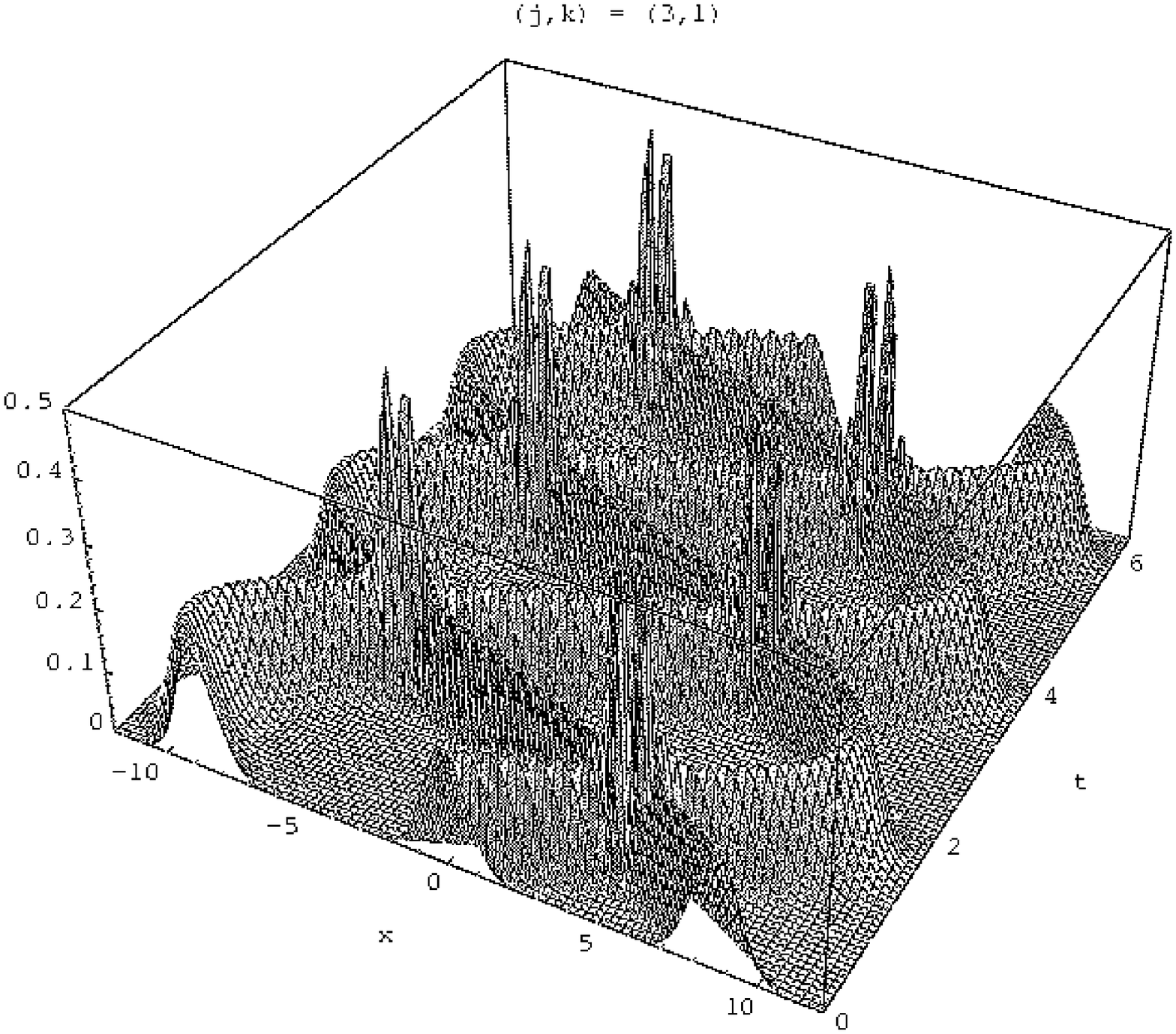,width=125mm,height=90mm}
  \caption{ The time evolution of $\rho_{(3,1)}(x,t)$ 
for the initial conditions $x_0 = 0$ and $p_0 = 10$. 
 \label{fig:hpcs3k=1}}
 \end{center}
 \end{figure}

\begin{figure}[p]
  \begin{center}
 \noindent    
\psfig{figure=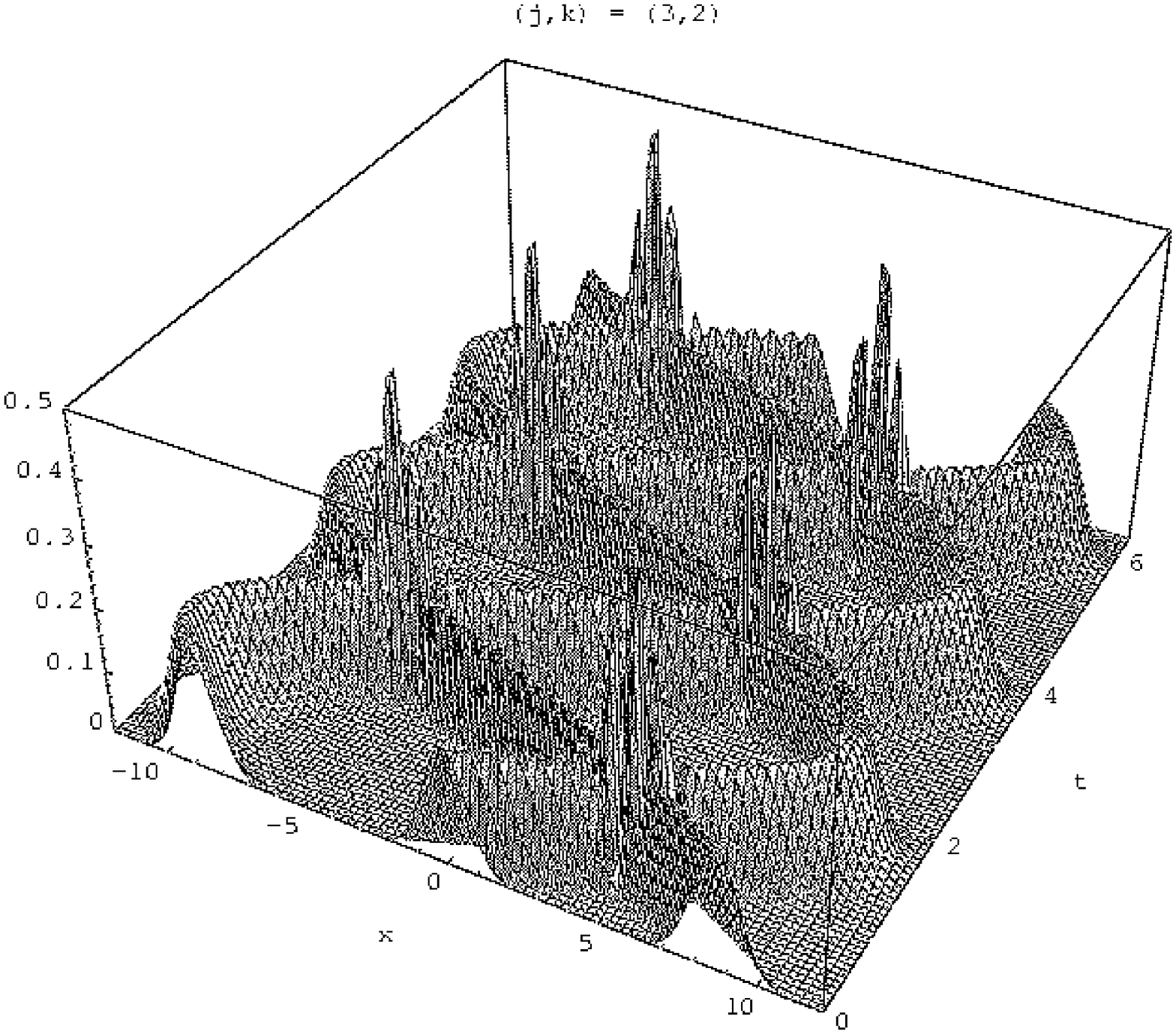,width=125mm,height=90mm}
  \caption{ The time evolution of $\rho_{(3,2)}(x,t)$ 
for the initial conditions $x_0 = 0$ and $p_0 = 10$. 
 \label{fig:hpcs3k=2}}
 \end{center}
\end{figure}

In Figures \ref{fig:hpcs3k=0}, \ref{fig:hpcs3k=1}, and \ref{fig:hpcs3k=2}  
we show the time-evolution of $\rho_{(3,k)}(x,t)$ for 
$k=0,1,2$, with initial conditions $x_0 = 0$ and $p_0 = 10$.  
With these initial conditions, the interference patters are more peaked 
than in Figs. \ref{fig:eocsfig1} and \ref{fig:eocsfig2}.  All three 
figures show interference patterns that do  not exactly look like 
either an even or odd interference.  This is not surprising, since the 
basic phase angle is $2\pi/3$.  However, Figure 
\ref{fig:hpcs3k=0} [$(j,k)=(3,0)$] and 
Figure \ref{fig:hpcs3k=1}  
[$(j,k)=(3,1)$] have interferences 
closer to odd ones, while Figure \ref{fig:hpcs3k=2}  
[$(j,k)=(3,2)$] has interferences closer to even ones.

%************

%**************

%************

%**************

%************

%**************

%*********************j=4 CS*******************
\newpage
\subsection{The Case $j=4$}

Now consider the $j=4$ states.  Define
\begin{eqnarray}
N_{(4,0)} &=& \cosh{A} + \cos{A},  ~~~~~~~~~
%     \label{N40} \\
N_{(4,1)}  =  \sinh{A} + \sin{A},
     \label{N41} \\
N_{(4,2)} &=& \cosh{A} - \cos{A},  ~~~~~~~~~
%     \label{N42}  \\
N_{(4,3)}  =  \sinh{A} - \sin{A},
     \label{N43}  \\
Z_1&=&\exp\left[-\lfrac{1}{2}(x -x_0)^{2}\right] e^{ip_0(x-x_0)},
    \label{Z1}  \\
Z_2&=&\exp\left[-\lfrac{1}{2}(x-p_0)^{2}\right]  e^{-ix_0(x-p_0)}, 
     \label{Z2}  \\
Z_3&=&\exp\left[-\lfrac{1}{2}(x+x_0)^{2}\right]  e^{-ip_0(x+x_0)},   
\label{Z3} \\
Z_4&=&\exp\left[-\lfrac{1}{2}(x+p_0)^{2}\right]  e^{ix_0(x+p_0)},   
\label{Z4}  \\
\theta_{1,2} &=& x(x_0+p_0) - p_0x_0,~~~~
\theta_{1,3} = 2xp_0,  ~~~~
\theta_{1,4} = x(p_0-x_0) - x_0p_0,  \\
\theta_{2,3} &=& x(p_0-x_0) + x_0p_0,~~~~
\theta_{2,4} = 2xx_0,~~~~
\theta_{3,4} = -x(x_0+p_0) - x_0p_0.
\end{eqnarray}

The states $\psi_{(4,k)}(x)$ are then
\begin{eqnarray}
\psi_{(4,0)}(x) &=& \frac{e^{A/2}}{2^{3/2} \pi^{1/4} N_{(4,0)}^{1/2}}
     \left[Z_1 + Z_2 + Z_3 + Z_4\right],
\label{psi40} \\
\psi_{(4,1)}(x) &=& \frac{e^{A/2}}{2^{3/2} \pi^{1/4} N_{(4,1)}^{1/2}}
     \left[Z_1 + i Z_2 - Z_3 -i Z_4\right],
\label{psi41} \\
\psi_{(4,2)}(x) &=& \frac{e^{A/2}}{2^{3/2} \pi^{1/4} N_{(4,2)}^{1/2}}
     \left[Z_1 - Z_2 + Z_3 - Z_4\right],
\label{psi42} \\
\psi_{(4,3)}(x) &=& \frac{e^{A/2}}{2^{3/2} \pi^{1/4} N_{(4,3)}^{1/2}}
     \left[Z_1 -iZ_2 - Z_3 +i Z_4\right]. 
\label{psi43} 
\end{eqnarray}
The probability densities $\rho_{(4,k)}(x)$ are
\begin{eqnarray}
\rho_{(4,0)}(x) &=& \frac{e^{A}}{8 \pi^{1/2} N_{(4,0)}}
     \left\{|Z_1|^2 + |Z_2|^2 + |Z_3|^2 + |Z_4|^2 
\right.    \nonumber \\  
&~&~~~
     +2 \cos{\theta_{1,2}}|Z_1Z_2|+2 \cos{\theta_{1,3}}|Z_1Z_3|
     +2 \cos{\theta_{1,4}}|Z_1Z_4|  \nonumber \\
&~&~~~\left.    
     +2 \cos{\theta_{2,3}}|Z_2Z_3|+2 \cos{\theta_{2,4}}|Z_2Z_4|
     +2 \cos{\theta_{3,4}}|Z_3Z_4|
\right\},
\label{rho40} \\
\rho_{(4,1)}(x) &=& \frac{e^{A}}{8 \pi^{1/2} N_{(4,1)}}
     \left\{|Z_1|^2 + |Z_2|^2 + |Z_3|^2 + |Z_4|^2 
\right.    \nonumber \\  
&~&~~~
     +2 \sin{\theta_{1,2}}|Z_1Z_2|-2 \cos{\theta_{1,3}}|Z_1Z_3|
     -2 \sin{\theta_{1,4}}|Z_1Z_4|  \nonumber \\
&~&~~~\left.    
     +2 \sin{\theta_{2,3}}|Z_2Z_3|-2 \cos{\theta_{2,4}}|Z_2Z_4|
     +2 \sin{\theta_{3,4}}|Z_3Z_4|
\right\},
\label{rho41} \\
\rho_{(4,2)}(x) &=& \frac{e^{A}}{8 \pi^{1/2} N_{(4,2)}}
     \left\{|Z_1|^2 + |Z_2|^2 + |Z_3|^2 + |Z_4|^2 
\right.    \nonumber \\  
&~&~~~
     -2 \cos{\theta_{1,2}}|Z_1Z_2|+2 \cos{\theta_{1,3}}|Z_1Z_3|
     -2 \cos{\theta_{1,4}}|Z_1Z_4|  \nonumber \\
&~&~~~\left.    
     -2 \cos{\theta_{2,3}}|Z_2Z_3|+2 \cos{\theta_{2,4}}|Z_2Z_4|
     -2 \cos{\theta_{3,4}}|Z_3Z_4|
\right\},
\label{rho42} \\
\rho_{(4,3)}(x) &=& \frac{e^{A}}{8 \pi^{1/2} N_{(4,3)}}
     \left\{|Z_1|^2 + |Z_2|^2 + |Z_3|^2 + |Z_4|^2 
\right.    \nonumber \\  
&~&~~~
     -2 \sin{\theta_{1,2}}|Z_1Z_2|-2 \cos{\theta_{1,3}}|Z_1Z_3|
     +2 \sin{\theta_{1,4}}|Z_1Z_4|  \nonumber \\
&~&~~~\left.    
     -2 \sin{\theta_{2,3}}|Z_2Z_3|-2 \cos{\theta_{2,4}}|Z_2Z_4|
     -2 \sin{\theta_{3,4}}|Z_3Z_4|
\right\}.
\label{rho43} 
\end{eqnarray}

\begin{figure}[p]
 \begin{center}
\noindent    
\psfig{figure=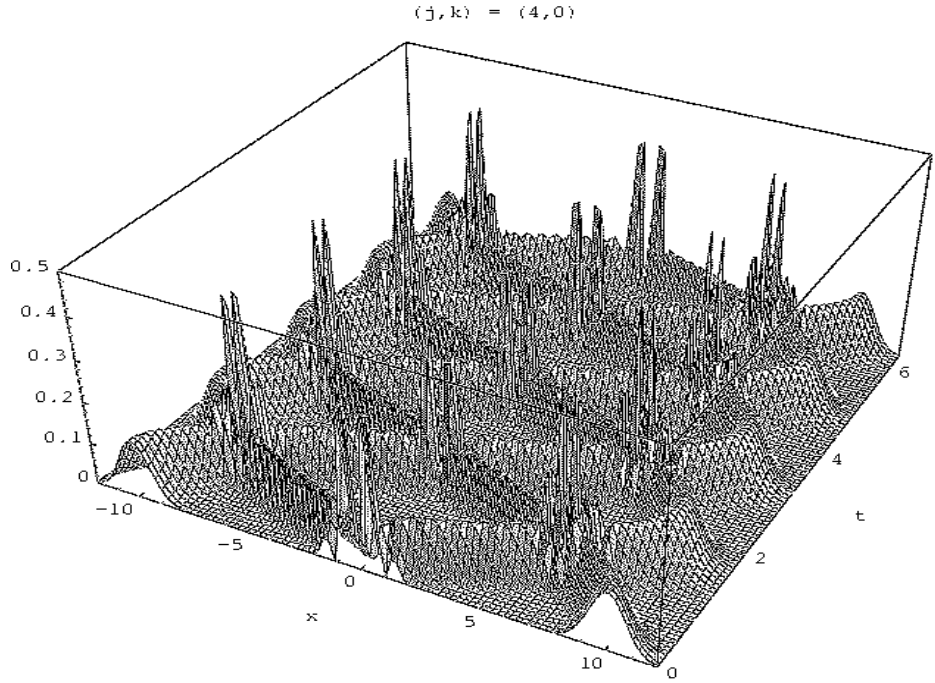,width=125mm,height=90mm}
  \caption{ The time evolution of $\rho_{(4,0)}(x,t)$ 
for the initial conditions $x_0 = 0$ and $p_0 = 10$. 
 \label{fig:hpcs4k=0}}
 \end{center}
 \end{figure}

\begin{figure}[p]
 \begin{center}
\noindent    
\psfig{figure=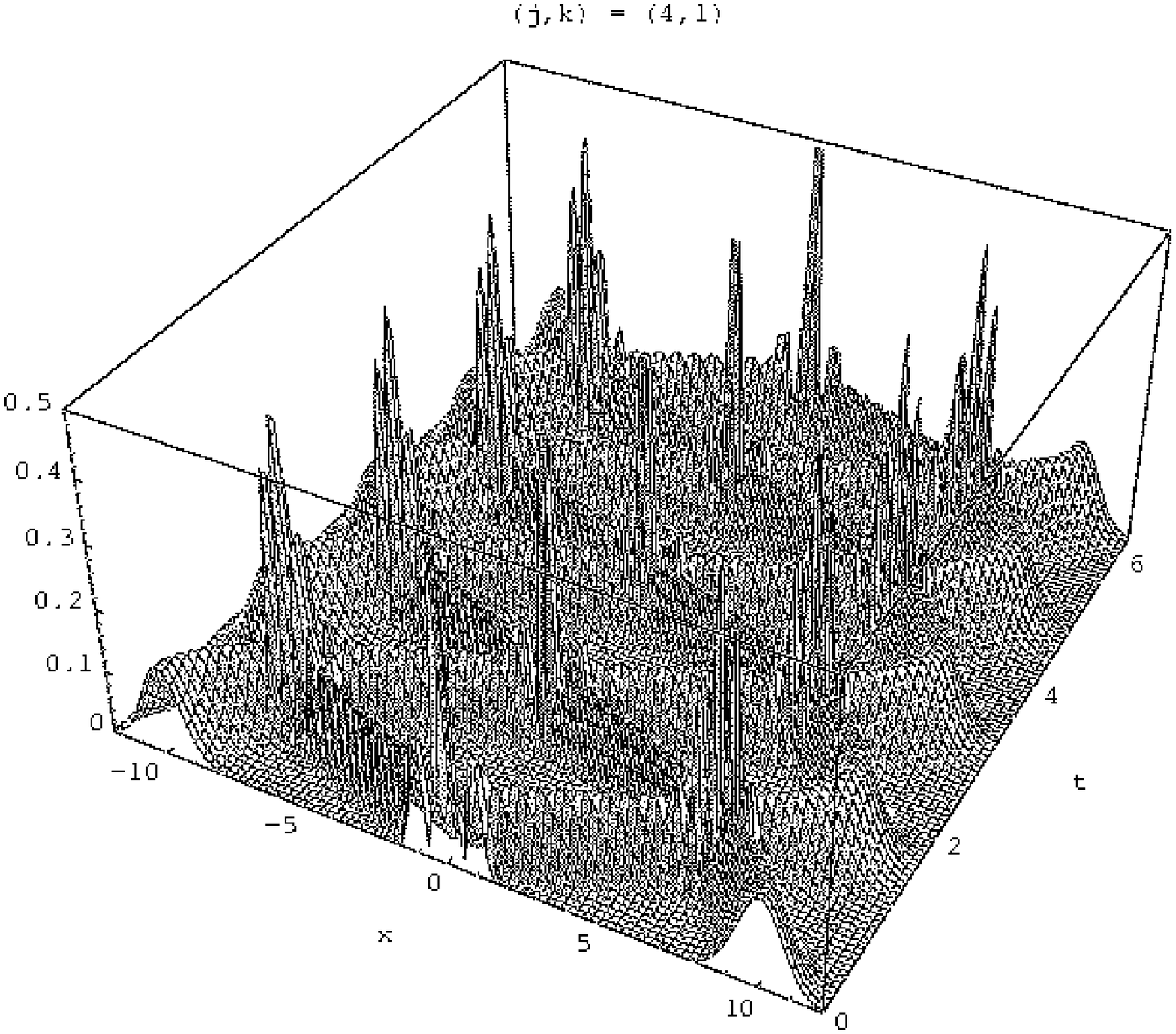,width=125mm,height=90mm}
  \caption{ The time evolution of $\rho_{(4,1)}(x,t)$ 
for the initial conditions $x_0 = 0$ and $p_0 = 10$. 
 \label{fig:hpcs4k=1}}
 \end{center}
 \end{figure}

\begin{figure}[p]
 \begin{center}
\noindent    
\psfig{figure=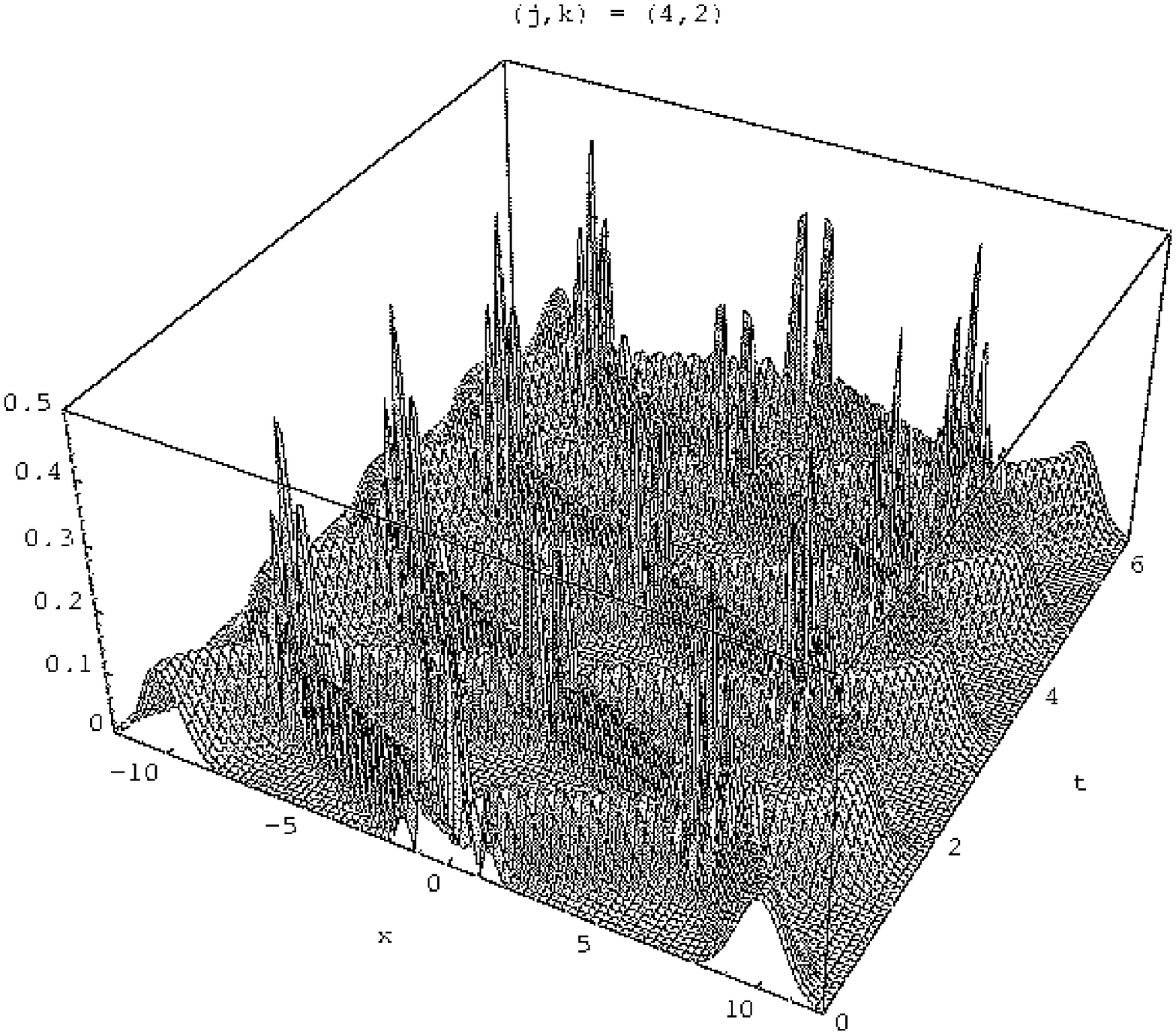,width=125mm,height=90mm}
  \caption{ The time evolution of $\rho_{(4,2)}(x,t)$ 
for the initial conditions $x_0 = 0$ and $p_0 = 10$. 
 \label{fig:hpcs4k=2}}
 \end{center}
 \end{figure}

\begin{figure}[p]
 \begin{center}
\noindent    
\psfig{figure=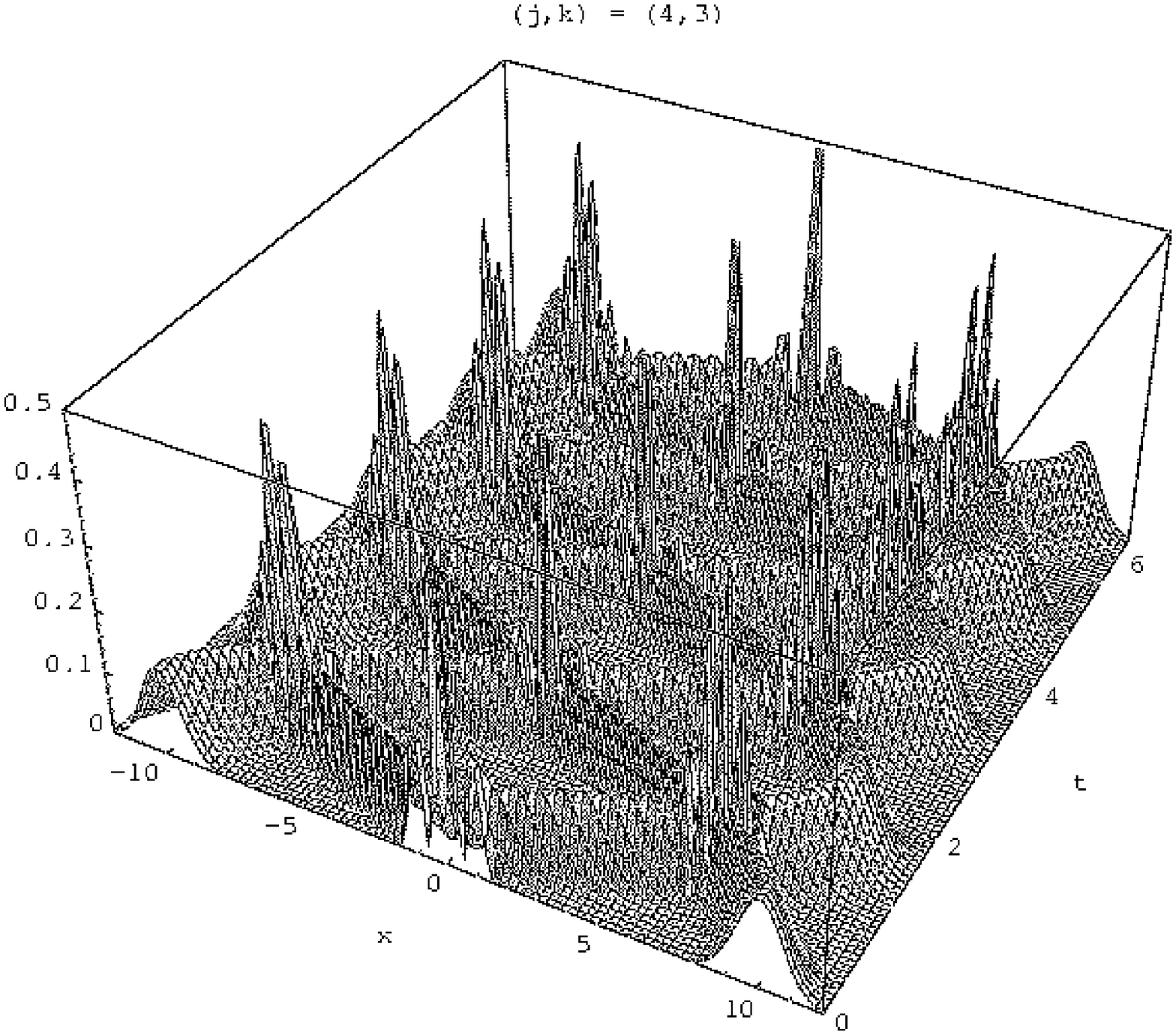,width=125mm,height=90mm}
  \caption{ The time evolution of $\rho_{(4,3)}(x,t)$ 
for the initial conditions $x_0 = 0$ and $p_0 = 10$. 
 \label{fig:hpcs4k=3}}
 \end{center}
 \end{figure}

In Figures  \ref{fig:hpcs4k=0} to  \ref{fig:hpcs4k=3}, 
we show the time-evolution of $\rho_{(4,k)}(x,t)$ for 
$k=0,1,2,3$, with initial conditions $x_0 = 0$ and $p_0 = 10$.  
Again, with these initial conditions the interference patters are more 
peaked than in Figs. \ref{fig:eocsfig1} and \ref{fig:eocsfig2}.  
In this example, all four 
figures show interference patterns that resemble  
either  even or odd interferences.  This is still not surprising since 
this time the 
basic phase angle is $\pi/2$.  Figure \ref{fig:hpcs4k=0} 
[$(j,k)=(4,0)$] has interferences 
that appear odd.  Figure \ref{fig:hpcs4k=1}  
[$(j,k)=(4,1)$] has central interferences that are even, 
but outer interferences that are closer to odd.   Figure \ref{fig:hpcs4k=2} 
[$(j,k)=(4,2)$] has central interferences that are odd, but outer
interferences that are even.  Finally, Figure \ref{fig:hpcs4k=3} 
[$(j,k)=(4,3)$] has interferences that all appear even.

%************

%**************

%************

%**************

%************

%**************

%************

%**************

%*****************************************************************
\newpage
\section{``Effective" Displacement-Operator Squeezed \\ States.}

The DOSS for the harmonic oscillator are well-known:
\begin{equation}
S(z)D(\alpha)|0\rangle=|\beta\rangle, ~~~
S(z)aS^{-1}(z)|\beta\rangle
         =(\mu a + \nu a^{\dagger})|\beta\rangle =\beta|\beta\rangle,
\label{ss1}
\ee
\be
 \mu = \cosh r,~~~~~~ \nu = -e^{i\phi}\sinh r,
~~~~~~\beta = [(\mu+\nu)x_0 + i(\mu-\nu)p_0]/\sqrt{2},
\label{ss2}
\ee
\be
|\beta\rangle \rightarrow  
     \left[\frac{(\mu+\nu)}{\pi^{1/2}(\mu-\nu)}\right]^{1/2}
   \exp\left[-\frac{(x-x_0)^2}{2}\frac{(\mu+\nu)}{(\mu-\nu)} +ip_0x\right], 
       \label{ss}
\end{equation}
where the $su(1,1)$ squeeze operator is
\be
S(z) = \exp{[za^{\da 2}/2 - z^*a^2/2]}, ~~~~~~z =re^{i\phi}, \label{opS}
\ee

When one tries to extend this definition to high-$j$ HPSS, 
one runs into problems. 
Similar to what was said for HPCS  with $j > 2$, 
there is no group-theoretic method to define 
higher-power squeeze operators  for $j>1$ \cite{nagel2, doss}.  
For $j>1$,  $a^{2j}$ and $(a^{\da})^{2j}$
do not form part of a closed algebra.   

However, there is an ``effective" displacement-operator
{\it ansatz} that can be used \cite{doss}.  First, one applies the 
ordinary squeeze operator, $S(z)$, 
to $a^j$.  By then inserting $I=S^{-1}(z)S(z)$
after each $a$, one finds
\be
[S(z)aS^{-1}(z)]^j |\bar{\beta};j,k\rangle = 
\bar{\beta}^j |\bar{\beta};j,k\rangle, ~~~~~~~    
|\bar{\beta};j,k\rangle =S(z) |\alpha;j,k\rangle.
\label{dossaj}
\ee

These squeezed states $|\bar{\beta};j,k\rangle$ 
are eigenvalues of the operator
\be
[S(z)aS^{-1}(z)]^j = (\bar{\mu} a + \bar{\nu} a^{\da})^j,
           ~~~~~~~ |\bar{\mu}|^2 - |\bar{\nu}|^2 =1.
\ee
Similar to before, these states are a subset of those states which 
satisfy equality of the Heisenberg uncertainty relation 
\begin{equation}
(\Delta {\cal{X}}_j)^2 (\Delta {\cal{P}}_j)^2 \ge \lfrac{1}{4}
               |\langle[{\cal{X}}_j,{\cal{P}}_j]\rangle|^2.   
\label{ehur}
\end{equation}
where
\begin{equation}
{\cal X}_j = \frac{(\bar{\mu} a + \bar{\nu} a^{\da})^j + 
         [(\bar{\mu} a + \bar{\nu} a^{\da})^j]^{\dagger}}
                   {\sqrt{2}},
     ~~~~~~~~~{\cal{P}}_j = \frac{(\bar{\mu} a + \bar{\nu} a^{\da})^j - 
         [(\bar{\mu} a + \bar{\nu} a^{\da})^j]^{\dagger}}
                   {i\sqrt{2}}.
\label{exp}
\end{equation}
[See. Eq. (\ref{mueq})].  The subset is defined by the additional 
restriction that $\Delta {\cal{X}}_j /\Delta {\cal{P}}_j=1$.

The $j$ Gaussians of a $(j,k)$ HPCS  are thereby transformed into 
$j$ squeezed Gaussians, similar to the ordinary squeezed-state
Gaussian of Eq. (\ref{ss}).  
Then, for example, the even- and odd-like interferences in Figures
\ref{fig:hpcs3k=0}-\ref{fig:hpcs4k=3} 
will become even- and odd-like squeezed interferences.  They 
will be  similar  
to those shown in Ref. \cite{eoss} for the even and odd squeezed states 
with $z$ real (with $s = e^r$),  
\begin{equation}
\psi_{ss\pm} =\left[\pi^{1/2}2s(1\pm e^{-x_0^2/s^2-p_0^2s^2})\right]^{-1/2}
    \left[e^{-(x-x_0)^2/(2s^2)+ip_0x} \pm
          e^{-(x+x_0)^2/(2s^2)-ip_0x}\right]~.
\end{equation}

%**********************************************************

\section{Ladder-Operator/Minimum-Uncertainty Squeezed States.}

When $j>1$, the ladder-operator (LO)/minimum-uncertainty  
(MU) method has its own problem. 
However, it is a technical problem, rather than one of principle. 
To calculate the squeezed states in closed form, becomes increasingly 
difficult as $j$ increases.  

This LO method has as its defining equations 
\begin{eqnarray} 
[\mu^j a^j + \nu^j a^{\da j}]|\beta;j,k\rangle&=&{\beta}^j|\beta;j,k\rangle,
             ~~~~~~ |{\mu}^j|^2 - |{\nu}^j|^2 =1,   \label{loss}  \\
|\beta;j,k\rangle &=& {\cal N}^{-1}(j,k)
              \sum_{n=0}^{\infty} c_{n}(j,k) |nj+k\rangle,~~~~~k < j,
\end{eqnarray}
where ${\cal N}(j,k)$ is the normalization constant.  This time 
the equivalent MU method yields that these states satisfy 
equality for (the more general) {\it Schr\"odinger} uncertainty 
relation \cite{schur}
\begin{equation}
(\Delta {X_j})^2 (\Delta {P_j})^2 \ge 
            \lfrac{1}{4}|\langle[X_j,P_j]\rangle|^2  +
  \lfrac{1}{4}\langle\{(X_j-\bar{X}_j), (P_j-\bar{P}_j)\}\rangle^2,
\label{sur}
\end{equation}
$\{,\}$ being the anticommutator.  By comparing 
Eqs. (\ref{hur}) and Eq. (\ref{sur}), one can then appreciate that the 
equation whose wave-function solution  minimizes this uncertainty 
relation is of the form of Eq. (\ref{mueq}), except that $B$ can now 
be complex and there is no restriction on $\Delta X/\Delta P$.  
(For the ordinary SS, this means the Gaussians need not have the 
width of the ground state and the squeeze can be complex.) 

Eq.  (\ref{loss}) yields a three-term recursion relation \cite{bender} 
among the  coefficients  $c_{n+1}(j,k)$, $c_{n}(j,k)$, and  
$c_{n-1}(j,k)$.  It is 
\be
{\mu^j}\left[\frac{[(n+1)j+k]!}{(nj+k)!}\right]^{1/2} c_{n+1}(j,k) 
+ {\nu^j}\left[\frac{[nj+k]!}{[(n-1)j+k]!}\right]^{1/2}c_{n-1}(j,k)
=  {\beta^j}c_{n}(j,k),
\label{crecur}
\ee
with boundary conditions determined by
\be
 c_{1}(j,k) =\left(\frac{\beta}{\mu}\right)^j
                   \left[\frac{k!}{(j+k)!}\right]^{1/2}c_{0}(j,k).
\ee
For $j>2$ this recursion relation has not been completely solved.  
But Nagel has studied it \cite{nagel2} from a Jacobi-matrix formulation. 
We proceed from his viewpoint.

Define 
\be
B = \frac{\beta}{\mu}, ~~~~~ R = \left(\frac{\nu \mu}{\beta^2}\right)^j,
 ~~~~~c_{n}(j,k) = \frac{b_{n}(j,k)B^{nj+k}}{[(nj+k)!]^{1/2}}.
\ee
We now have the recursion relation
\be 
b_{n+2}(j,k) = b_{n+1}(j,k) -R~b_{n}(j,k)\frac{((n+1)j+k)!}{(nj+k)!}, 
                          \label{brecurs}
\ee
with boundary conditions (yet to be normalized) of 
\be
b_0(j,k) = 1, ~~~~~~~ b_1(j,k) = 1.  \label{bzrecurs}
\ee

Introduce the notation 
\be
T_{n}(j,k) = \frac{(nj+k)!}{((n-1)j+k)!} = 
((n-1)j+k+1)_j,
\ee
where $(\alpha)_N=\Gamma(\alpha +N)/\Gamma(\alpha)$ is the Pochhammer 
symbol.  The solutions for $b_n(j,k)$ are
\begin{eqnarray}
b_0(j,k)      &=& b_1(j,k) = 1, \label{b0}  \\
b_2(j,k) &=& 1 - R\left[T_1\right],  \\
b_3(j,k) &=& 1 - R\left[T_1+T_2\right],  \\
b_4(j,k) &=& 1 - R\left[T_1+T_2+T_3\right]  
                      +R^2\left[T_1 T_3\right],  \\
b_5(j,k) &=& 1 - R\left[T_1+T_2+T_3 +T_4\right]  
        +R^2\left[T_1 T_3 +T_1 T_4+T_2 T_4 \right], \\
b_6(j,k) &=& 1 - R\left[T_1+T_2+T_3 +T_4+T_5\right]  
                 +R^2\left[T_1 T_3 +T_1 T_4+T_1 T_5 \right. \nonumber  \\
         &~&~~~~~~~~ \left.+T_2 T_4 +T_2 T_5++T_3 T_5\right]
                     -R^3\left[T_1 T_3 T_5\right].  \label{b6}
\end{eqnarray}
Higher-$n$ solutions continue with the same pattern.

The pattern is, first of all, a power series in $(-R)^{[\lfrac{n}{2}]}$.  
The factor multiplying 
$(-R)$ is the sum of all possible $T$'s, up to $T_{n-1}$.  The factor 
multiplying 
$(-R)^2$ is the sum of all possible products of two $T$'s, that differ by 
order of at least two, up to the quantity $T_{n-1}$.  
The factor multiplying $(-R)^3$ is the 
sum of all products of  three $T$'s, each differing by order of
at least two from the others, up to the quantity $T_{n-1}$, and so on.  
Symbolically, this is 
\begin{eqnarray}
b_n(j,k) &=& \sum_{t=0}^{\left[\lfrac{n}{2}\right]}(-R)^t
          \sum_{\forall v=1}^{n-1}~\prod_{i=1}^{t}T_{v_i}(j,k)~, 
               \nonumber \\
             &~&~~~~~~~~~
    1 \le v_1 \le  v_2-2\le \dots \le v_{n-2}-2 \le v_{n-1} \le n-1.
\end{eqnarray}
Using these  $b$'s, 
\be
|\beta;j,k\rangle = {\cal N}^{-1}(j,k)\sum_{n=0}^{\infty} b_n(j,k)
         \frac{B^{nj+k}}{[(nj+k)!]^{1/2}} |nj+k\rangle,
\ee
\be
{\cal N}^{2}(j,k) = \sum_{n=0}^{\infty} 
         \frac{(B^*B)^{nj+k}}{[(nj+k)!]} b_n(j,k)^*b_n(j,k).
\ee
Converting to the number-state wave functions 
\be
\psi_n(x) = \frac{\exp[-\frac{1}{2}x^2]H_n(x)}
              {[\pi^{1/2}2^n n!]^{1/2}},
\ee
we have
\be
|\beta;j,k\rangle \rightarrow \frac{\exp[-\frac{1}{2}x^2]}
             {{\cal N}(j,k)\pi^{1/4}}
           \sum_{n=0}^{\infty} b_n(j,k) 
        \left(\frac{B}{2^{1/2}}\right)^{nj+k}\frac{H_{nj+k}(x)}{(nj+k)!},
   \label{sumss}
\ee

If the $b_n(j,k)$'s were a power in $[f(j,k)R]^{[n/2]}$
(which their values are suggestive of), the sum in Eq. (\ref{sumss}) 
would be of the form 
of the generating function $G$ of Eq. (\ref{Gjk}).   One would again have 
$j$ orthogonal wave functions ($k=0,1,\dots,j-1$), each containing 
$j$ Gaussians (differently-squeezed than before), distributed 
evenly around the phase-space circle.  

Finally, as observed by Nagel \cite{nagel2}, one can demonstrate 
normalizability of the states.  Note that, for large $n$, the 
recursion relation (\ref{brecurs}) is dominated by  
\be
b_{n+2}(j,k) \approx - R ~b_n(j,k)T_{n+1}(j,k).
\ee
Then, the even and odd coefficients are decoupled and the highest-order 
in $R$ contribution is dominant.  (This pattern is seen in the specific 
$b$'s we gave above.)  Taking for definiteness the even-$n$ case, use 
Stirling's approximation to evaluate $T_{n+1}(j,k)$.  One ends up with 
an exponent of a sum which one changes to an exponent of an integral.  
When one is finished evaluating, one finds 
\be
{\cal N}^{2}(j,k) \rightarrow Const. \sum_{v=0}^{\infty} \left(
         \frac{|\nu^2|^j}{|\mu^2|^j}\right)^{v/2}.
\ee
\noindent From Eq.  (\ref{loss}), this is a convergent geometric series. 
(Similarly for the odd-$n$ case.)

%************************************************************

\section{Special-Case Solutions for the LO/MU Squeezed States.} 

We start by  considering the special case $(j,k)=(1,0)$.  Here, 
the LO/MU-SS are  
identical to the DO-SS of Eqs. (\ref{ss1})-(\ref{ss}): 
\be
 \mu = \cosh r,~~~~~~ \nu = -e^{i\phi}\sinh r,
~~~~~~\beta = [(\mu+\nu)x_0 + i(\mu-\nu)p_0]/\sqrt{2},
\label{ss2N}
\ee
\be
|\beta;1,0\rangle \rightarrow \psi_{ss}(x) =  
     \left[\frac{(\mu+\nu)}{\pi^{1/2}(\mu-\nu)}\right]^{1/2}
   \exp\left[-\frac{(x-x_0)^2}{2}\frac{(\mu+\nu)}{(\mu-\nu)} +ip_0x\right]. 
       \label{ssN}
\ee

The decomposition into number states is straight forward, and well known. 
However, it is enlightening to show how it fits into our general scheme. 
With the aid of Eqs. (\ref{ss1})-(\ref{ss}) 
the $c_n$'s can be obtained from $c_n=(\psi_n,\psi_{ss})$.  
This yields the $b_n(1,0)$'s, which are
\begin{eqnarray}
b_n(1,0) &=& \left({R}\over{2}\right)^{n/2} H_n\left[(2R)^{-1/2}\right]
           = R^{n/2} He_n[R^{-1/2}]   \label{bH}  \\   
   &=& \sum_{t=0}^{\left[\lfrac{n}{2}\right]} (-R)^t
      \left[\frac{n!}{2^t (t!)[(n-2t)!]}\right]  \label{bSum}  \\
   &=& \left(-\frac{R}{2}\right)^{\left[\frac{n}{2}\right]}
       \frac{n!}{\left[\frac{n}{2}\right]!}  
       ~{_1F_1}\left(-\left[\frac{n}{2}\right];\frac{2+(-1)^{n+1}}{2};
             \frac{1}{2R}\right),  \label{bF}
\end{eqnarray}
where $_1F_1$ is the confluent hypergeometric function.
This agrees with the special $(j,k)=(1,0)$ examples
of our formulae (\ref{b0})-(\ref{b6}).  
Further, putting the first equality of Eq. (\ref{bH}) and $T_n(1,0)=n$
back into the recursion relation (\ref{brecurs}) yields the 
standard Hermite polynomial recursion relation, 
\be
H_{n+1}(x)= 2xH_n(x) - 2nH_{n-1}(x).
\ee

For $(j,k)=(2,0)$ and $(2,1)$, 
we have the ladder-operator even- and odd-squeezed states. 
The wave-function solutions 
are confluent hypergeometric 
functions \cite{nagel2,eoss,realmany,many,nagel1}:
\begin{eqnarray}
\psi_{(2,0)ss}(x) &=& N_{(2,0)} 
\exp{\left[-\frac{x^2}{2}(U+\sqrt{U^2-1})\right]}~\nonumber \\
~&~&~~~~~~~\times\,\, 
_1F_1\left(\left[\frac{1}{4}+\frac{B}{2\sqrt{U^2-1}}\right];~
\frac{1}{2};~x^2\sqrt{U^2-1}\right),  
\\
\psi_{(2,1)ss}(x)&=&N_{(2,1)}  
         x~ \exp{\left[-\frac{x^2}{2}(U+\sqrt{U^2-1})\right]}~
                    \nonumber   \\
      ~&~&~~~~~~~\times\,\,                 
_1F_1\left(\left[\frac{3}{4}+\frac{B}{2\sqrt{U^2-1}}\right];~
\frac{3}{2};~x^2\sqrt{U^2-1}\right),  \\
U &=& \frac{(\mu^2 - \nu^2)}{(\mu^2 + \nu^2)}, ~~~~~~
B = \frac{\beta^2}{(\mu^2 + \nu^2)}.
\end{eqnarray}
The shapes of the states so produced resemble the ``effective" DO-SS 
with their Gaussians \cite{eoss}.  

Nagel \cite{nagel2} discussed the decomposition into the 
number states $|2n+k\rangle$, $k=(0,1)$.  One obtains that the   
$c_{n}(2,k)$ are proportional to  Pollaczek 
polynomials  \cite{nagel2}, 
\be
P_n(x,\delta) \equiv i^n \sqrt{(2\delta)_n/n!}~{_2F_1}
           \left(-n,~(\delta+ix);~2\delta;~2\right).
\ee
When put in our notation, the results are
\begin{eqnarray}
b_n(2,0) &=&  i^n \left(\frac{1}{2}\right)_n 2^nR^{n/2}~{_2F_1}
\left(-n,~\left(\frac{1}{4}+\frac{i}{4R^{1/2}}\right);\right.
            \left. ~\frac{1}{2};~2\right),               \label{Fbn20}  \\
b_n(2,1) &=&  i^n \left(\frac{3}{2}\right)_n 2^nR^{n/2}~{_2F_1}
\left(-n,~\left(\frac{3}{4}+\frac{i}{4R^{1/2}}\right);\right.
            \left. ~\frac{3}{2};~2\right).               \label{Fbn21} 
\end{eqnarray}
Putting these $b_n$'s into Eq.  (\ref{brecurs}) 
reduces it to the Gauss contiguous relation
\be
0=(c-a){_2F_1}(a-1,b;c;z)+(2a-c-az+bz){_2F_1}(a,b;c;z) 
               +a(z-1){_2F_1}(a+1,b;c;z).
\ee

%**********************************************************************

\section{Experimental Realizations}

The simplest higher-power coherent states, 
the even- and odd-coherent states, are commonly called 
``Schr\"odinger Cat States,'' since they are the mathematical 
realizations of Schr\"odinger's {\it gedanken} cat that is 
simultaneously dead and alive.  Although this ``cat'' was for 
decades a matter of heated epistemological debate, with the 
advent of modern quantum optics such states have now been demonstrated 
in two different physical systems.    

Wineland's group \cite{wine} was able, with much effort, to entangle 
$^9{\rm Be}^+$ ions in a trap, producing even- and odd- coherent states. 
The method starts with the ion in its vibrational ground state. 
Then,  with a $\pi/2$ laser pulse the hyperfine levels are mixed.
A different kicking laser excites only the upper hyperfine level into an 
energetic coherent state. Both sets of internal states are then swapped 
by a $\pi/2$ laser pulse.  Finally, the motionless component, is 
excited by  by a new kicking pulse, yielding a mixed state.  
Mind you, in general they produce two wave packets with differing relative
phases.  The relative phases must be adjusted to $0$ and $\pi$ to yield  
the orthonormal even- and odd-states.  

The second system, studied by Haroche's group \cite{haroche}, 
a rubidium atom is prepared in a mixture of circular $n=50$ and $n=51$ 
Rydberg states.  This atom is then sent through a high-Q cavity with 
a few coherently produced photons in it.  The traversing atom's two 
states  shift the phases of the photons differently, thus producing 
entangled field/atom states.  By re-mixing the the Rydberg states 
after the atom leaves the cavity, EPR states can be produced.  
[See Ref. \cite{haroche2} for a popular account of both these systems.]  

Both systems appear amenable, in principle, to  extensions producing 
3-power or 4-power coherent states.  However, this would be complicated 
to actually achieve in practice. %carry through.  
Even so, such extensions would be of 
interest.  For example,  as Gerry has emphasized \cite{gerry}, 
the 3-power states arise in models of trapping.  
In the two-channel model of Jyotsna and Agarwal \cite{ja}, there are 
certain trapping states that involve CS that are eigenstates of 
$a^3$.  These are explicitly the 3-power coherent  states.

Additionally, all these states could be, again in principle, 
squeezed \cite{wine2, opts}.  

%************************************************************

\subsection*{Acknowledgements}

We thank Bengt Nagel for illuminating communications.   We also thank 
George A. Baker, Jr. for discussions on normalizability.  
MMN acknowledges the support of the United States Department of 
Energy and the Alexander von Humboldt Foundation.  DRT acknowledges
a grant from the Natural Sciences and Engineering Research Council 
of Canada.

%**Bibliography**************************************************

%****************************************************************
%***************************************************************

\end{document}